\documentclass[english]{article}
\usepackage{geometry}
\usepackage{float}
\usepackage{mathrsfs}
\usepackage{amsmath}
\usepackage{amssymb}
\usepackage{graphicx}
\usepackage{amsfonts}
\usepackage{bm}
\usepackage{subfigure}
\usepackage{epsfig}
\usepackage{xcolor}
\usepackage{setspace}
\usepackage{caption}
\usepackage{multirow}

\makeatletter

\begin{document}

\begin{center}

{\Large \textbf{A sequential algorithm for fast fitting of Dirichlet process mixture models}}

\bigskip

BY DAVID J. NOTT$^{1}$, XIAOLE ZHANG$^{2}$, CHRISTOPHER YAU$^{2}$ \& AJAY JASRA$^{1}$

{\footnotesize $^{1}$Department of Statistics \& Applied Probability,
National University of Singapore, Singapore, 117546, SG.}\\
{\footnotesize E-Mail:\,}\texttt{\emph{\footnotesize standj@nus.edu.sg, staja@nus.edu.sg}}\\
{\footnotesize $^{2}$Department of Mathematics,
Imperial College London, London, SW7 2AZ, UK.}\\
{\footnotesize E-Mail:\,}\texttt{\emph{\footnotesize x.zhang11@imperial.ac.uk, c.yau@imperial.ac.uk}}

\end{center}

\begin{abstract}
In this article we propose an improvement on the sequential updating and greedy search (SUGS) algorithm \cite{wang} for fast fitting of Dirichlet process mixture models. The SUGS algorithm provides a means for very fast approximate Bayesian inference for mixture data which is particularly of use when data sets are so large that many standard Markov chain Monte Carlo (MCMC) algorithms cannot be applied efficiently, or take a prohibitively long time to converge. In particular, these ideas are used to initially interrogate the data, and to refine models such that one can potentially apply exact data analysis later on. SUGS relies upon sequentially allocating data to clusters and proceeding with an update of the posterior on the subsequent allocations and parameters which assumes this allocation is correct. Our modification softens this approach, by providing a probability distribution over allocations, with a similar computational cost; this approach has an interpretation as a variational Bayes procedure and hence we term it  variational SUGS (VSUGS). It is shown in simulated examples that VSUGS can out-perform, in terms of density estimation and classification, the original SUGS algorithm in many scenarios. In addition, we present a data analysis for flow cytometry data,
and SNP data via a three-class dirichlet process mixture model  illustrating the apparent improvement over SUGS.
\end{abstract}

\section{Introduction}

The demands of fitting models to large data-sets have exploded over the last decade. Increasingly complex data sets are available, which has placed demands on statisticians to develop realistic models to represent these data. Inevitably, for many classes of models, this places a further emphasis on being able to fit such models accurately and in a reasonable time-frame. 

In this article, we consider fast Bayesian statistical inference for Dirichlet process mixture (DPM) models \cite{antoniak,lo}. This particular class of models have proven to be popular in the literature as a tool for both clustering and density estimation
and there are a wide variety of elegant MCMC and sequential Monte Carlo algorithms; see e.g.~\cite{maceachern1,ulker}. Such algorithms provide exact inference from DPM models, but can be very computationally demanding when trying to analyze extremely large data-sets and even more, exact statistical inference from mixtures is notoriously difficult; see \cite{jasra}. As mentioned above, this issue often leads to researchers resorting to approximate inference to browse or interrogate the data, so as to refine model specifications for an exact analysis; one particular important and interesting method in this direction is the SUGS algorithm.

The SUGS algorithm is a procedure for fast approximate fitting of DPM models. It relies on an approximation of the exact posterior distribution on the allocation of data to components and the component specific parameters, by sequentially adding data-points to the model. These data are allocated to a given mixture component and this allocation is frozen and taken as the truth when updating the posterior for new data points. The method can be sensitive to the ordering of the data, but \cite{wang} provide procedures to select this ordering in a systematic way. In this paper, we reinterpret the SUGS algorithm within a variational Bayes framework. This allows one to derive different approximations of the posterior distribution.

In particular, our interpretation does not mean that one needs to allocate data to a cluster and, instead, provides a probability distribution on these allocations; this is done at a minor increase in computational cost. The advantage of this generalization, which we call VSUGS, is apparent when one fits the approximation to data whose components are close in some sense. In this scenario, we have consistently found (and as illustrated in Section \ref{sec:simos}) that VSUGS outperforms SUGS in a variety senses; this is important as, when the mixture components are highly separated, such initial browsing or interrogation of the data is less important. Moreover, our variational approximation provides a lower-bound on the log-marginal likelihood; we empirically find that this can be used as a technique for model selection (e.g.~selecting the order in which the data arrive) which did not seemingly work well in \cite{wang}.

This article is structured as follows. 
We begin with a motivating example in Section \ref{sec:motivating}.
In Section \ref{sec:dpm} we give a basic summary of DPM models. In Section \ref{sec:sugs} we discuss SUGS and in Section \ref{sec:vsugs} our generalization VSUGS. In Section \ref{sec:simos} we give numerical examples; both a simulation study and real data analyses associated to flow cytometry data and SNP data via a three-class dirichlet process mixture model. In Section \ref{sec:summ} we conclude the article, discussing avenues for future work.

\section{Motivating Application}\label{sec:motivating}

A motivating application for us is the problem of genotyping single nucleotide polymorphisms (SNPs) from SNP genotyping microarray data \cite{giannoul}. Figure \ref{fig:genotyping_example} illustrates an example dataset. The statistical problem is to characterise the three genotype classes $AA$, $AB$ and $BB$ and classify each data point into one of these three classes. This is a straightforward three-way classification problem that can be approached using hierarchical mixture modelling where, for example, the class-conditional densities are modelled using multivariate Normal or Student $t$-distributions. Although, these models work well in practice, it is clear that the class-conditional densities are not Normal (or Student). We can obtain increased accuracy through the use of semi-parametric models for the class-conditional densities using Dirichlet Process Mixtures. However, the size of the data sets presents a massive challenge for this type of modelling approach. For a single experiment (individual), modern genotyping microarray produces 300,000-5,000,000 two-dimensional measurements. Each study may consist of hundreds to thousands of individuals. The data sizes here prohibit the use of Monte Carlo inference and motivate approximate approaches that are able to scale to the size of problems encountered.

\section{Dirichlet process mixture model}\label{sec:dpm}

Consider a Dirichlet process mixture model  of the form, for $i\in\mathbb{N}$:
\begin{equation}
Y_i|\tilde{\theta}_i \stackrel{\textrm{i.i.d.}}{\sim} P(\cdot|\tilde{\theta}_i)\;\;\;\;\;\;\tilde{\theta}_i\sim P\;\;\;\;\;\;P\sim DP(\alpha P_0) \label{dpmmodel}
\end{equation}
where $Y_i\in\mathsf{Y}\subseteq\mathbb{R}^{d_y}$, $\tilde{\theta}_i\in\Theta\subseteq\mathbb{R}^{d_{\theta}}$ are observation specific parameters,
$P(\cdot|\tilde{\theta}_i)$ is a conditional probability which admits a density $p(\cdot|\tilde{\theta}_i)$ w.r.t.~a single dominating
$\sigma-$finite measure for each $\tilde{\theta}_i$ (which is often Lebesgue), 
 $P$ is
an unknown mixing distribution, and $P\sim DP(\alpha P_0)$ indicates that the prior
for $P$ is a Dirichlet process \cite{ferguson} with precision parameter $\alpha\in\mathbb{R}_+$
and base measure $P_0$.  $P_0$ is known and we also
consider $\alpha$ to be fixed. We remark that, in connection to subsequent methodology to be presented, \cite{wang} consider a way of handling
unknown $\alpha$ which can also be used in all the extensions we consider but
for simplicity we do not consider this below.  

A well known property of the Dirichlet process is that a distribution
drawn from it will put all its mass on a countable
set of points.  Following the notation of \cite{wang} we will write
$\theta=\{\theta_j\}_{j=1}^\infty$ for the set of distinct values
in the sequence $\tilde{\theta}=\{\tilde{\theta}_j\}_{j=1}^\infty$ where points in $\theta$ 
are labelled according to their order of appearance in $\tilde{\theta}$.  Next, let $\delta_i=j$
if $\tilde{\theta}_i=\theta_j$ and write $\delta=\{\delta_j\}_{j=1}^\infty$.  Using
the P\'{o}lya urn characterization of the Dirichlet process \cite{blackwell}
we can rewrite (\ref{dpmmodel}) in the form
\begin{equation*}
Y_i|\delta_i,\theta_{\delta_i} \stackrel{\textrm{i.i.d.}}{\sim} P(\cdot|\theta_{\delta_i}) \;\;\;\;\;\; 
 p(\delta,\theta)=p(\delta)p(\theta) 
\end{equation*}
where $p(\theta)=\prod_{j=1}^\infty p_0(\theta_j)$, 
$p_0(\cdot)$ is the probability density associated to $P_0$ and
$p(\delta)$   is as follows, (writing $\delta_{1:i}=(\delta_1,...,\delta_i)$, with the convention $\delta_{1:0}$ is the null vector), for $i\in \{2,3,\dots\}$
\begin{eqnarray}
p(\delta_i=j|\delta_{1:i-1}) & = & \left\{\begin{array}{cc}
  \frac{n_j^{(i)}}{\alpha+i-1} & j\in\{1,\dots,n_i\} \\
  \frac{\alpha}{\alpha+i-1} & j=n_i+1 
\end{array}\right.  \label{polyaurn1}
\end{eqnarray}
where $p(\delta_1=1)=1$, $n_j^{(i)}=\textrm{Card}(\{\delta_m:\delta_m=j, 1\leq m\leq i-1\})$, 
 and $n_i=n_i(\delta_{1:i-1})$ is
the maximum value in $\delta_{1:i-1}$ (i.e.~the number of components ``seen" in the
data up to to time $i-1$).  This representation of the model where the unknown measure
$P$ is integrated out is important for many Monte Carlo sampling schemes for fitting DP
mixtures \cite{bush,escobar1,maceachern,maceachern1}.  

Later we will work with a truncated Dirichlet process mixture model, which is often convenient
for computations \cite{ishwaran}.  
Such truncations are based on the stick breaking representation of the Dirichlet process
\cite{sethu}.  Suppose we limit the number of distinct values appearing in the
sequence $\{\tilde{\theta}_j\}_{j=1}^\infty$ to an upper truncation limit $T>1$. 
Then generalizing the P\'{o}lya urn representation we can consider the truncated
Dirichlet process with $p(\delta,\theta)=p(\delta)p(\theta)$ where now
$p(\theta)=\prod_{j=1}^T p_0(\theta_j)$ and $p(\delta)$ is defined recursively by
$p(\delta_1=1)=1$ and similarly to (\ref{polyaurn1}) (see, for example, \cite[Section 2.2.2]{ishwaran}) for $i\in\{2,3,\dots\}$
\begin{eqnarray}
p(\delta_i=j|\delta_{1:i-1}) & = & \left\{\begin{array}{cc}
  \frac{n_j^{(i)}+\alpha/T}{\alpha+i-1} & j\in\{1,\dots,n_i\} \\
  \frac{\alpha(1-n_i/T)}{\alpha+i-1} & j=n_i+1 
\end{array}\right.  \label{polyaurn2}
\end{eqnarray}

We note that truncations have also been used in the context of 
variational approximations for Dirichlet process mixtures \cite{blei} although
these authors consider the truncation point as a variational parameter without truncating
the original model.  The algorithm of \cite{blei}, although related to ours, 
is not a sequential algorithm however - here we are interested in very fast sequential
algorithms related to the SUGS method of \cite{wang}.  \cite{wang}
compare their approach with a variety of other fast computational methodologies for Dirichlet
process mixtures, and show that their algorithm is competitive with other fast approximation
methodologies.  In this work we focus only on comparing our new approach with
the original SUGS algorithm, and refer the reader to \cite{wang} for information
about the relative performance of alternative approximations to SUGS.  

\section{The SUGS algorithm}\label{sec:sugs}

 SUGS is a recursive algorithm that takes at time $i-1$ an estimate
$\hat{\delta}_{1:i-1}$ of $\delta_{1:i-1}$ and an approximation of $p(\theta|y_{1:i-1})$ and
produces an estimate $\hat{\delta}_{1:i}$ of $\delta_{1:i}$ and an
approximation of $p(\theta|y_{1:i})$.  To start the recursion we use
$\hat{\delta}_1=1$ and $p(\theta|y_1)=p(\theta_1|y_1,\hat{\delta}_1)\prod_{j>1} p_0(\theta_j).$
Here and in what follows a term $\hat{\delta}_i$ in the conditioning means $\delta_i=\hat{\delta}_i$ and
similarly for $\hat{\delta}_{1:i}$.  Suppose at time $i-1$ we have an estimate $\hat{\delta}_{1:i-1}$
of $\delta_{1:i-1}$.  Consider the posterior distribution 
$p(\delta_i,\theta|y_{1:i})$ and approximate this by
$p(\delta_i,\theta|y_{1:i},\hat{\delta}_{1:i-1})$. That is to say, we initially consider our approximation
of $p(\delta_i,\theta|y_{1:i})$, $\widehat{p}_i(\delta_i,\theta|y_{1:i})$,  as
\begin{eqnarray*}
 \widehat{p}_i(\delta_i,\theta|y_{1:i})  & \propto & p(\delta_i,\theta|y_{1:i-1},\hat{\delta}_{1:i-1}) p(y_i|\theta_{\delta_i}) \\
   & = & p(\delta_i|\hat{\delta}_{1:i-1})
   \left\{\prod_{j\in\mathbb{N}} p(\theta_j|\hat{\delta}_{1:i-1},y_{1:i-1})\right\}
   p(y_i|\theta_{\delta_i})     
\end{eqnarray*}
Using this approximation and integrating out $\theta$, 
\begin{eqnarray*}
\widehat{p}_i(\delta_i|y_{1:i}) & \propto & p(\delta_i|\hat{\delta}_{1:i-1}) \int p(\theta_{\delta_i}|\hat{\delta}_{1:i-1},y_{1:i-1})p(y_i|\theta_{\delta_i})d\theta_{\delta_i}.
\end{eqnarray*}
The SUGS algorithm sets:
\begin{equation}
\hat{\delta}_i := \textrm{argmax}_{\delta_i\in \{1,...,n_i(\hat{\delta}_{1:i-1})+1\}}\Big[\widehat{p}(\delta_i|y_{1:i})\Big]
\label{eq:hat_delta}
\end{equation}
 and then one replaces
$$
p(\delta_i|\hat{\delta}_{1:i-1}) p(\theta_{\delta_i}|\hat{\delta}_{1:i-1},y_{1:i-1})p(y_i|\theta_{\delta_i})
$$
by
$$
\mathbb{I}_{\{\hat{\delta}_i\}}(\delta_i) p(\theta_{\delta_i}|\hat{\delta}_{1:i-1},y_{1:i-1})p(y_i|\theta_{\delta_i})
$$
to form the approximation:
\begin{equation}
\widehat{p}_i(\delta_i,\theta|y_{1:i})= \mathbb{I}_{\{\hat{\delta}_i\}}(\delta_i)
\left\{\prod_{j\neq \hat{\delta}_i}p(\theta_j|\hat{\delta}_{1:i-1},y_{1:i-1})\right\}
p(\theta_{\hat{\delta}_i}|\hat{\delta}_{1:i},y_{1:i}).
\label{eq:sugs_approx_post}
\end{equation}
 In this approximation
the components $\theta_j$ are independent for different $j$, the posterior for
$\theta_j$ for $j\neq \hat{\delta}_i$ is unchanged from time $i-1$, and 
the posterior for $\theta_{\hat{\delta}_i}$ is updated by assuming that $\delta_i=\hat{\delta}_i$
so that $y_i$ represents an observation from this mixture component.  If the mixture
components are from the exponential family and conjugate priors are used, the densities
such as $p(\theta_j|\hat{\delta}_{1:i},y_{1:i})$ can be calculated in closed
form and sufficient statistics are updated recursively, leading to a very efficient update.  

\cite{wang} consider a number of further innovations in their algorithm.  
First, since the fitting algorithm is sequential and there is a dependence of the
fit on the ordering of the data, they suggest running their algorithm for different
random orderings and then choosing the best according to a pseudo-likelihood criterion.  
Secondly, for model comparison they suggest the approximation
$$
\widehat{p}(y_{1:i}) = p(y_{1:i}|\hat{\delta}_{1:i})
$$
and show that this crude approximation can be useful for tasks such as comparison of
parametric and nonparametric models.  Thirdly, they suggest a way of dealing with unknown
$\alpha$ in the Dirichlet process prior.  

\section{An improvement of the SUGS algorithm}\label{sec:vsugs}

Here we suggest a simple improvement of the SUGS algorithm which we call VSUGS (variational SUGS).  
We begin with a brief introduction to variational Bayes (VB) methods.  

\subsection{Variational Bayes}

Suppose we have a parameter $\xi\in\Xi\subseteq \mathbb{R}^{d_{\xi}}$ and data $y$, $p(\xi)$ is the prior density, $p(y|\xi)$ the likelihood
and $p(\xi|y)$ denotes the posterior density w.r.t.~Lebesgue measure (which we use for presentational purposes only).  In VB \cite{jordan,bishop} we split $\xi$ into blocks
$\xi=\xi_{1:k} = (\xi_1,...,\xi_k)$, $\xi_{j}\in\Xi_j$, with $\Xi_1\times\cdots\times\Xi_k = \Xi$
 and seek to find a good approximation
to $p(\xi_{1:k}|y)$ of the form 
$$
\widehat{q}(\xi_{1:k})=\prod_{j=1}^k q(\xi_j)
$$
where each $q(\xi_j)$ is a probability density w.r.t.~the appropriate dimensional Lebesgue measure.
 Given known probability densities for $q(\xi_j)$, $j\neq i$, the optimal
choice for $q(\xi_i)$ for minimizing the Kullback-Leibler divergence
\begin{equation}
 \textrm{KL}(\widehat{q}||p) := \int \log\bigg(\frac{\widehat{q}(\xi_{1:k})}{p(\xi_{1:k}|y)}\bigg)\widehat{q}(\xi_{1:k})d\xi_{1:k} \label{kld}
\end{equation}
is 
\begin{eqnarray}
 \widehat{q}(\xi_i) & \propto & \exp\{ \mathbb{E}_{-\widehat{q}(\xi_i)}[\log p(\Xi)p(y|\Xi)] \} 
\label{graddescent}
\end{eqnarray}
where $\mathbb{E}_{-\widehat{q}(\xi_i)}[\cdot]$ denotes expectation w.r.t.~$\prod_{j\neq i}\widehat{q}(\xi_j)$.  
Hence there is a gradient descent algorithm for minimizing \eqref{graddescent} based
on choosing initial values for the factors in $q(\eta)$ and then iteratively updating each
term according to \eqref{graddescent}.  Minimizing (\ref{kld}) is equivalent to
maximizing
\begin{eqnarray}
 L(q) & := & \int \log\bigg(\frac{p(\xi_{1:k})p(y|\xi_{1:k})}{\widehat{q}(\xi_{1:k})}\bigg)\widehat{q}(\xi_{1:k}) d\xi_{1:k} \label{lowerbd}
\end{eqnarray}
and \eqref{lowerbd} is a lower bound on the log marginal likelihood $\log p(y)$ where
$p(y)=\int p(\xi_{1:k})p(y|\xi_{1:k}) d\xi_{1:k}$.  $p(y)$ is a key quantity in Bayesian model
selection and the lower bound is tight, $L(q)=\log p(y)$, when $\widehat{q}(\xi_{1:k})=p(\xi_{1:k}|y)$.  Generally
$L(q)$ is often used as an approximation to $\log p(y)$ for model selection in the VB
framework.

\subsection{The VSUGS algorithm}

As we have seen in \eqref{eq:sugs_approx_post} the SUGS algorithm recursively approximates $p(\delta_{1:i},\theta|y_{1:i})$.  
Considering this in a variational framework, suppose we have an approximation to
the posterior $p(\delta_{1:i-1},\theta|y_{1:i-1})$ of the form
$$
\bigg\{\prod_{j=1}^{i-1} q_{i-1}(\delta_j)\bigg\}
\bigg\{\prod_{j\in\mathbb{N}} q_{i-1}(\theta_j)\bigg\}.
$$
In the above expression, we omit certain conditionings (as will become apparent below) to reduce the subsequent notational burdens.
We will suggest a way to update this approximation of
$p(\delta_{1:i},\theta|y_{1:i})$, using
variational ideas. The approximation will be of the form
$\{\prod_{j=1}^i q_{i}(\delta_j)\}\{\prod_{j\in\mathbb{N}} q_{i}(\theta_j)\}$.  
We start the recursion with
$\widehat{q}_{1}(\delta_1=1)=1$, $\widehat{q}_{1}(\theta_1)=p(\theta_1|y_1,\delta_1=1)$, 
$\widehat{q}_{1}(\theta_j)=p_0(\theta_j)$, $j\in\{2,3,\dots\}$.  

The idea is to make a particular fixed choice at time $i$ for $\widehat{q}_{i}(\delta_i)$ and not to revisit
that choice at future times.  That is, the solution to the (partial) variational optimization
at time $i-1$ is used to initialize the optimization at time $i$.  
The original SUGS algorithm chooses $\widehat{q}_{i}(\delta_i)=\mathbb{I}_{\{\hat{\delta}_i\}}(\delta_i)$
where $\hat{\delta}_i$ is defined in \eqref{eq:hat_delta}
 but it is
possible to make a better choice than this without sacrificing the attractive computational
properties of the original SUGS algorithm.  
In particular, at time $i$, we set
$\widehat{q}_{i}(\delta_j)=\widehat{q}_{i-1}(\delta_j)$, $j\in\{1,...,i-1\}$ and choose
\begin{eqnarray}
\widehat{q}_{i}(\delta_i=j) & = & q_{ij}
\int \widehat{q}_{i-1}(\theta_{\delta_i})
p(y_i|\theta_{\delta_i})d\theta_{\delta_i}  \label{qterm}
\end{eqnarray}
for $j\in\{1,\dots,i\wedge T\}$ where $T$ is a truncation point for the number of mixture components
and
$$
q_{ij}=\left\{\begin{array}{ll}
\frac{\sum_{k=1}^{(i-1)}\widehat{q}_{i-1}(\delta_k=j)+\alpha/T}{\alpha+i-1} & \mbox{$j\in\{1,\dots,(i-1)\wedge T\}$} \\
\frac{\alpha (1- [(i-1)\wedge T]/T)}{\alpha+i-1} & \mbox{$j= (i-1)\wedge T+1$}.
\end{array}\right.
$$  
$\widehat{q}_{i}(\delta_i=j)$ will be chosen as an approximation to $p(\delta_i|y_{1:i})$.  To provide some intuition
for this selection of  $\widehat{q}_{i}(\delta_i=j)$, we remark that
\begin{eqnarray}
 p(\delta_i|y_{1:i}) & \propto & p(\delta_i|y_{1:i-1})p(y_i|\delta_i,y_{1:i-1}) \nonumber \\
  & = & p(\delta_i|y_{1:i-1})\int p(y_i|\theta_{\delta_i})p(\theta_{\delta_i}|y_{1:i-1})d\theta_{\delta_i} \label{brule}
\end{eqnarray}
Next, note that 
$$
p(\delta_i|y_{1:i-1})=\mathbb{E}[p(\delta_i|y_{1:i-1},\delta_{1:i-1})|y_{1:i-1}]=\mathbb{E}[p(\delta_i|\delta_{1:i-1})|y_{1:i-1}]
$$
where the expectation is w.r.t.~$p(\delta_{1:i-1}|y_{1:i-1})$.  This suggests approximating
$p(\delta_i|y_{1:i-1})$ by taking the expectation in the above expression with respect to 
the variational posterior $\widehat{q}_{i-1}(\delta_{1:i-1})$.  Although this approximation is still not
easy to work with, if we we condition on $n_i= (i-1)\wedge T$ in (\ref{polyaurn2}) and then take
the expectation with respect to $\widehat{q}_{i-1}(\delta_{1:i-1})$, we get $q_{ij}$.  
So $q_{ij}$ is an approximation to $p(\delta_i|y_{1:i-1})$ in (\ref{brule}) and
the term 
$$
\int \widehat{q}_{i-1}(\theta_{\delta_i})
p(y_i|\delta_i,\theta_{\delta_i})d\theta_{\delta_i}
$$
in \eqref{qterm} simply approximates the integral in \eqref{brule} by replacing
$p(\theta_{\delta_i}|y_{1:i-1})$ with the variational posterior $\widehat{q}_{i-1}(\theta_{\delta_i})$.  
As noted earlier, the original SUGS algorithm can
be placed in our framework by using $\widehat{q}_{i}(\delta_i)=\mathbb{I}_{\{\hat{\delta}_i\}}(\delta_i)$, a 
``hard" rather than ``soft" allocation to clusters which tends to result in greater under-estimation
of uncertainty than in our approach.  

Next, using $\widehat{q}_{i-1}(\theta_j)$ as the prior for $\theta_j$ at time $i$ for processing
the data point $y_i$, the optimal choice for $\widehat{q}_i(\theta_j)$ is, via \eqref{graddescent} 
\begin{eqnarray}
  \widehat{q}_{i}(\theta_j) & \propto & \widehat{q}_{i-1}(\theta_j) 
   \exp\left(\mathbb{E}_{-\widehat{q}_{i}(\theta_j)}\left(\sum_{k=1}^{T_i} \mathbb{I}_{\{k\}}(\delta_i) \log\big(p(y_i|\Theta_k)\big)\right)\right) \nonumber \\
   & \propto & \widehat{q}_{i-1}(\theta_j)p(y_i|\theta_j)^{\widehat{q}_{i}(\delta_i=j)} \label{vathetaupdate}
\end{eqnarray}
where $\mathbb{E}_{-\widehat{q}_{i}(\theta_j)}$ denotes expectation w.r.t.~\{$\prod_{h=1}^i \widehat{q}_{i}(\delta_h)\}\{\prod_{h\neq j} \widehat{q}_{i}(\theta_h)\}$ and $T_i= i\wedge T$.
Again, with the choice $\widehat{q}_{i}(\delta_i)=\mathbb{I}_{\{\hat{\delta}_i\}}(\delta_i)$, exponential family
mixture components and conjugate priors this reduces
to the SUGS update.  If the mixture components are normal then the generalized update above can be done
in closed form -- we will give details of this below.  
Note the attractive form of the above update.  The likelihood contribution from the $i^{th}$ observation
is split among different mixture components $j$ according to the weight $\widehat{q}_{i}(\delta_i=j)$ rather
than assuming the most likely allocation as in the original SUGS algorithm.  
We can also use the variational lower bound (\ref{lowerbd})
to approximate the marginal likelihood more accurately than with $\log(p(y_{1:i}|\delta_{1:i}))$ in
the SUGS algorithm.  Details of this are given in the next section for the case of normal
mixture components. 

\subsection{VSUGS for DP mixtures of normals}\label{sec:dpm_normal}

\cite{wang} consider the case of DP mixtures of normals in detail.  
In this case, $\theta_i=(\mu_i,\zeta_i)$ where $\mu_i$ is the mean for the $i$th
component and $\zeta_i$ is the precision and we have
$Y_i|\delta_i=j,\theta_{\delta_i} \stackrel{\textrm{i.i.d.}}{\sim} \mathcal{N}(\mu_j,\zeta_j^{-1})$.  
They consider
a normal inverse-gamma prior for $\theta_j$, $p_0(\theta_j)=p_0(\mu_j|m,\nu \zeta_j^{-1})
p_0(\zeta_j|a,b)$ where $p_0(\mu_j|m,\nu \zeta_j^{-1})$ is a normal density with associated distribution
with mean $\rho$ and variance $\nu \zeta_j^{-1}$ and $\rho$ and $\nu$ are known hyperparameters, 
and $p_0(\zeta_j|a,b)$ is a gamma density with known parameters $a$ and $b$.  

Our VSUGS algorithm results in $\widehat{q}_{i}(\theta_j)$ being normal inverse-gamma also, 
$\widehat{q}_{i}(\theta_j)=\widehat{q}_{i}(\mu_j|\rho_j^{(i)},\nu_j^{(i)} \zeta_j^{-1}) \widehat{q}_{i}(\zeta_j|a_j^{(i)},b_j^{(i)})$
where $\widehat{q}_{i}(\mu_j|\rho_j^{(i)},\nu_j^{(i)} \zeta_j^{-1})$ is the normal density with mean
$\rho_j^{(i)}$ and variance $\nu_j^{(i)} \zeta_j^{-1}$, and $\widehat{q}_{i}(\zeta_j|a_j^{(i)},b_j^{(i)})$ is a gamma density
with parameters $a_j^{(i)}$ and $b_j^{(i)}$.  The parameters $\rho_j^{(i)}$, $\nu_j^{(i)}$, $a_j^{(i)}$
and $b_j^{(i)}$ are updated recursively by (c.f.~\cite[p.~204]{wang})
\begin{eqnarray*}
  \nu_j^{(i)} & = & \left\{(\nu_j^{(i-1)})^{-1}+\widehat{q}_{i}(\delta_i=j)\right\}^{-1} \\
  \rho_j^{(i)} & = & \nu_j^{(i)}\left\{\left\{\nu_j^{(i-1)}\right\}^{-1}\rho_j^{(i-1)}+\widehat{q}_{i}(\delta_i=j)y_i\right\} \\
  a_j^{(i)} & = & a_j^{(i-1)}+\frac{\widehat{q}_{i}(\delta_i=j)}{2} \\
  b_j^{(i)} & = & b_j^{(i-1)}+\frac{1}{2}\left\{\widehat{q}_{i}(\delta_i=j)y_i^2+\frac{{\rho_j^{(i)}}^2}{\nu_j^{(i)}}-\frac{{\rho_j^{(i-1)}}^2}{\nu_j^{(i-1)}}\right\}
\end{eqnarray*}
To calculate the terms $\widehat{q}_{i}(\delta_i=j)$ in the VSUGS algorithm, we also need to evaluate the
integral
$$
\int \widehat{q}_{i-1}(\theta_j) p(y_i|\theta_j)d\theta_j.
$$
In the normal case with the priors we have chosen this integral evaluates to a $t$ density, 
$t_{2a_j^{(i-1)}}(y_i;\rho_j^{(i-1)},b_j^{(i-1)}/$ $a_j^{(i-1)}(\nu_j^{(i-1)}+1))$ where
$t_d(y;m,s^2)$ denotes a $t$ density for $y$ with $d$ degrees of freedom, location parameter $m$ and
scale parameter $s$.  

An approximate variational lower bound on $\log p(y_{1:i})$ can also be computed recursively.  
We can think of the posterior at stage $i-1$ as the prior to be updated by the likelihood
contribution for the $i$th observation:  
$$p(\delta_i,\theta|y_{1:i})\propto p(\delta_i,\theta|y_{1:i-1})p(y_i|\theta_{\delta_i}).$$
Approximating $p(\delta_i|y_{1:i-1})$ by $q_{ij}$
as we did previously and approximating $p(\theta|y_{1:i-1})$ by $\prod_{j=1}^{T_i} \widehat{q}_{i-1}(\theta_j)$ 
and calculating the lower bound (\ref{lowerbd}) using these priors for the likelihood
contribution $p(y_i|\theta_{\delta_i})$ gives
\begin{eqnarray*}
L(q) & = & \sum_{j=1}^T \left\{(a_j^{(i)}-a_j^{(i-1)})\psi(a_j^{(i)}) 
-\log \big(\Gamma(a_j^{(i)})\big)+\log\big(\Gamma(a_j^{(i-1)})\big)+ a_j^{(i-1)}\left(\log\big( b_j^{(i)}\big)-\log \big(b_j^{(i-1)}\big)\right) \right. \\
& & \left.
+a_j^{(i)}\frac{(b_j^{(i-1)}-b_j^{(i)})}{b_j^{(i)}}+
\frac{(\rho_j^{(i)}-\rho_j^{(i-1)})^2}{2\nu_j^{(i-1)}}\frac{a_j^{(i)}}{b_j^{(i)}}
+\frac{1}{2}\left(\frac{\nu_j^{(i)}}{\nu_j^{(i-1)}}-1-\log\bigg(\frac{\nu_j^{(i)}}{\nu_j^{(i-1)}}\bigg)\right)\right\} \\
 & & +\sum_{j=1}^T\widehat{q}_{i}(\delta_i=j)\left\{\frac{1}{2}\psi(a_j^{(i)})-\frac{1}{2}\log\big( b_j^{(i)}\big)
 -\frac{1}{2}\log \big(2\pi\big)-\frac{1}{2}\left\{\nu_j^{(i)}+(y_j-\rho_j^{(i)})^2\frac{b_j^{(i)}}{a_j^{(i)}}\right\}\right\} \\
 & & -\sum_{j=1}^{T_i} \widehat{q}_{i}(\delta_i=j)\log\big(\widehat{q}_{i}(\delta_i=j)\big) + \sum_{j=1}^{T_i}\widehat{q}_{i}(\delta_i=j) \log\big(q_{ij}\big)
\end{eqnarray*}
where $\psi(\cdot)$ denotes the digamma function.  

\section{Numerical examples}\label{sec:simos}

\subsection{Density Estimation}

\begin{figure}[!h]
\centering
	\includegraphics[width=\textwidth]{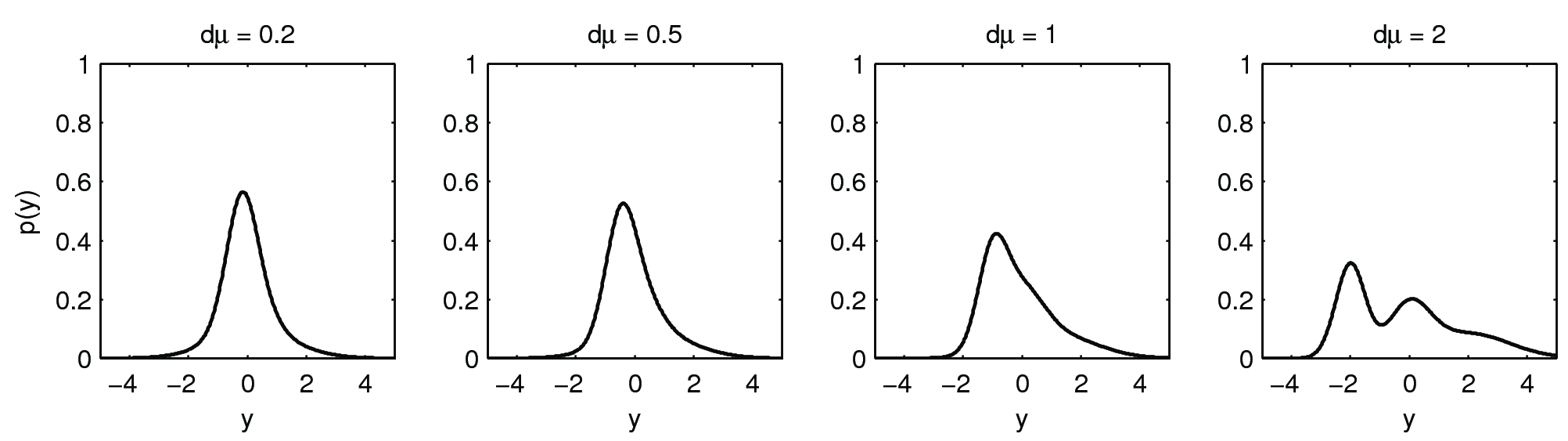}
	\caption{Example probability densities for four simulated mixture datasets with $d\mu = 0.2, 0.5, 1.0$ and $2.0$.}
	\label{fig:mixtures}
\end{figure}

{\it Setup.} We generated 100 data sets for each setting and all results are averaged over that for each data set. 
$$
Y_i \stackrel{\textrm{i.i.d.}}{\sim} \frac{2}{5}\mathcal{N}(-d\mu,0.25) + \frac{3}{10}\mathcal{N}(0,0.5) + \frac{3}{10}\mathcal{N}(d\mu,2), i = 1, \dots, N ,
$$
for $0 \leq d\mu \leq 5$ with $N = 500$. Examples are shown in Figure \ref{fig:mixtures}.

We compared the relative errors of SUGS to VSUGS under different values of $\alpha$, i.e. $\alpha\in\{0.1,~0.5,~1,~5,~10\dots{}45,~50\}$. Throughout $T=200$ for VSUGS. We used 50 different (but random) orderings of the data and chose the ordering with the maximal variational lower-bound for VSUGS in Section \ref{sec:dpm_normal} and the best ordering for SUGS as in \cite{wang}. We also used a standard Collapsed Gibbs Sampling method \cite{neal} for posterior inference on some of the datasets for comparison. To assess the performance in density estimation we compute the value of
\begin{equation}
	e = \sum_{j=1}^N (\hat{f}(y_j)-f(y_j))^2
\label{eq:quality_measure}
\end{equation}
where $\hat{f}(y_j)$, $f(y_j)$, are the estimated (predictive) and true density for the data, evaluated at data-point $y_j$ and examine the relative errors.

\begin{figure}[!h]
\centering
	\includegraphics[width=\textwidth]{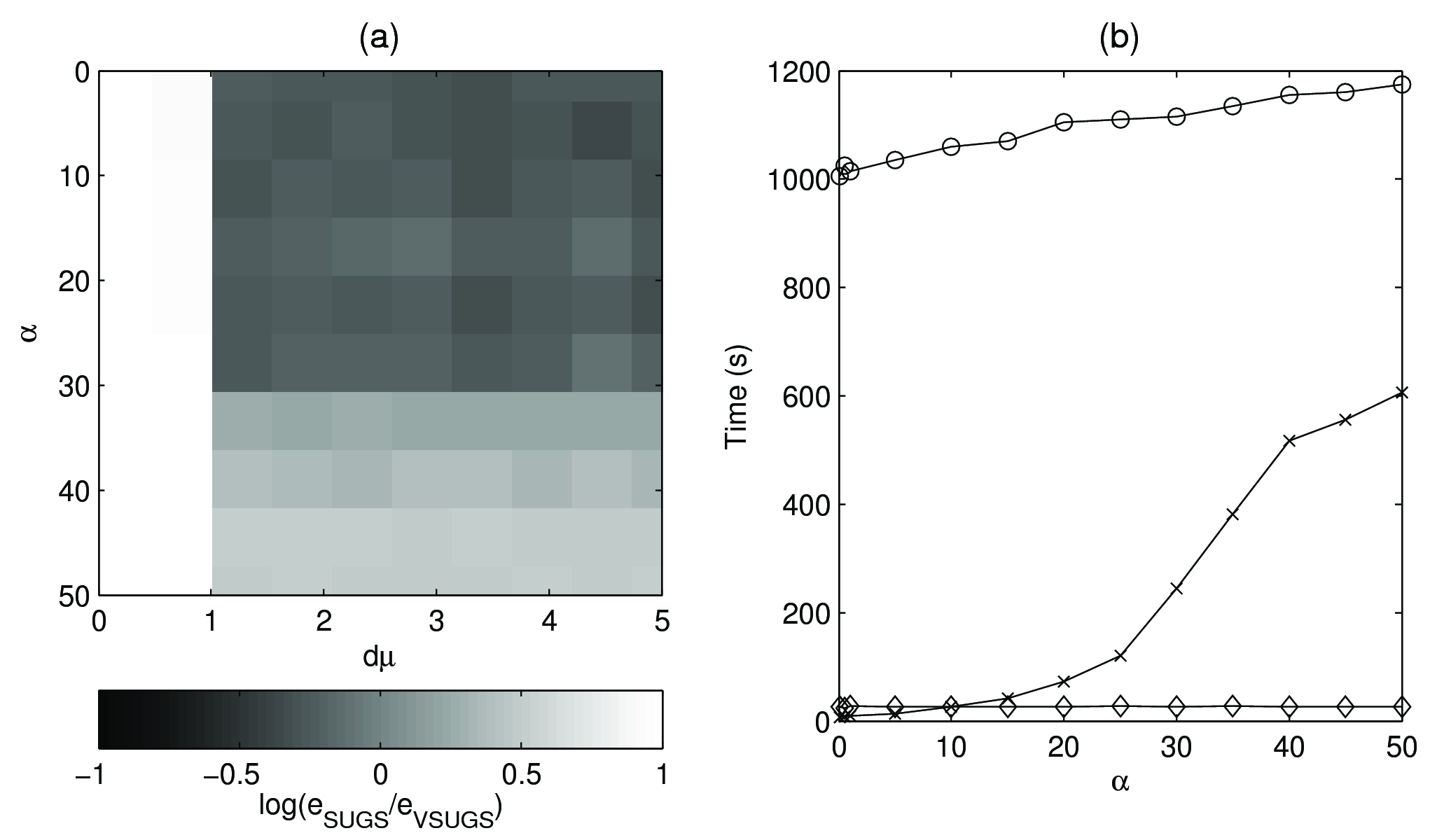}
	\caption{(a) (Log) relative error of SUGS to VSUGS as a function of $(d\mu, \alpha)$. (b) Computational times for Gibbs Sampling, SUGS and VSUGS. Results are averaged over 100 data sets.}
	\label{fig:relative_error}
\end{figure}

\begin{figure}[!h]
\centering
	\includegraphics[width=\textwidth]{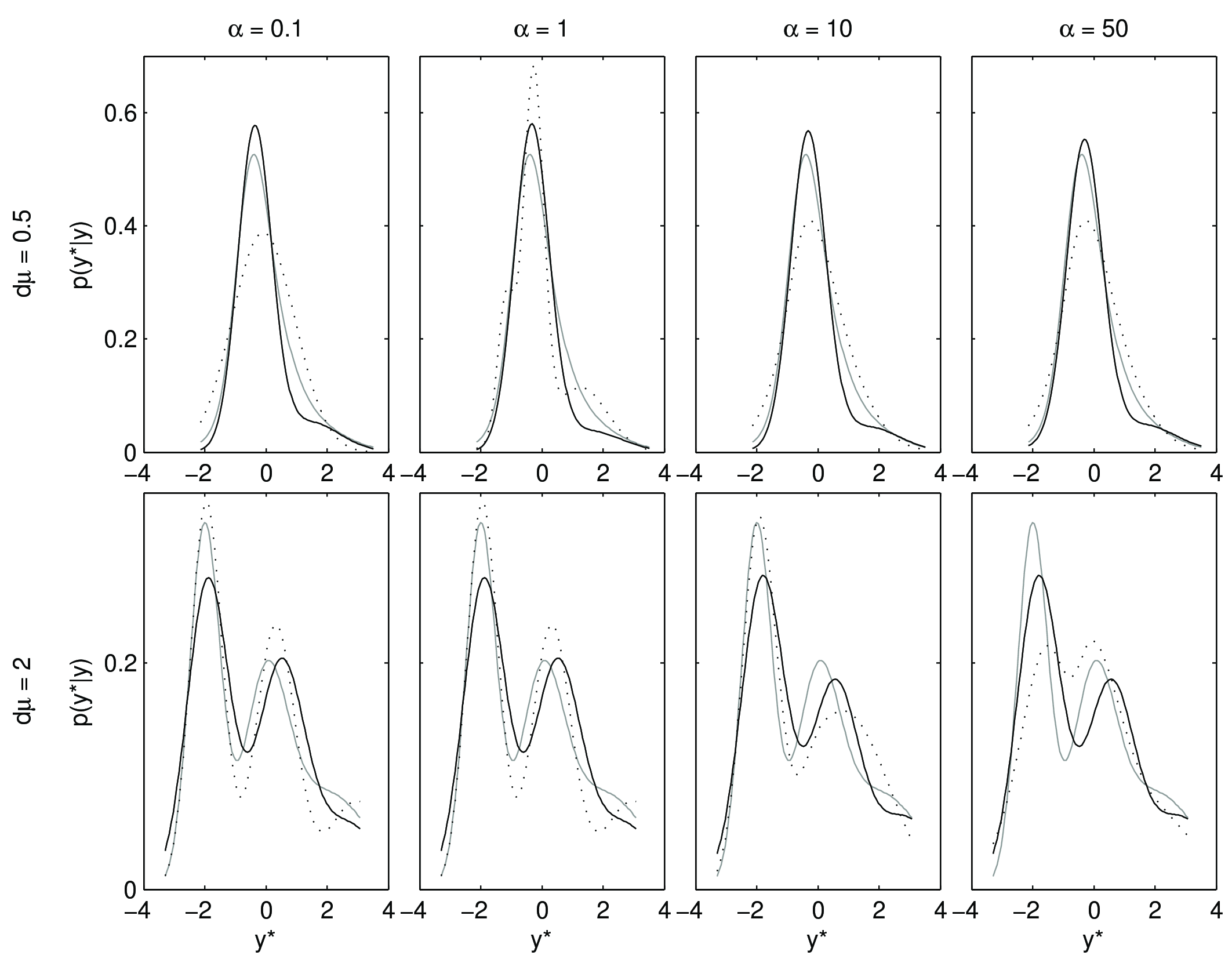}
\caption{Examples of fitted predictive densities (Gray) Truth, (Dotted) SUGS and (black) VSUGS.}\label{fig:dens_examples}
\end{figure}

{\it Results.} Figure \ref{fig:dens_examples} shows example predictive density estimates from SUGS and VSUGS. Figure \ref{fig:relative_error}(a) shows that for large values of $\alpha$ and closely spaced clusters $d\mu < 1$, VSUGS provides more accurate density estimates than SUGS. However, for $d\mu > 1$ and $\alpha < 20$, i.e.~well-separated clusters, the density estimates from SUGS are relatively more accurate. 

The computation time for VSUGS is constant for given truncation level $T$ as we use a fixed maximum number of mixture components. In contrast, the computation time required for SUGS is variable and depends both on the data set and the order in which the data is processed. Figure \ref{fig:relative_error}(b) considers the computation burden for the two methods. In particular, for large values of $\alpha$ and more mixture components, SUGS can be computationally quite demanding due to the excessive numbers of mixture components that are realised. Whilst in practice, one might estimate $\alpha$, this value is not known and hence SUGS could both be significantly less accurate and computationally more expensive in many situations.

We compared the SUGS and VSUGS predictive densities with those obtained from Collapsed Gibbs Sampling, we considered the case $d\mu = 0.2, \alpha = 0.1$ and show results in Table \ref{tab:comparison_to_gibbs} for different data sizes $N$ and the truncation parameter $T$. Using Collapsed Gibbs Sampling as a ``gold standard", we find that VSUGS consistently provides better predictive density estimates. Example computational times for $N=500$ were $4$ seconds for SUGS, $12$ seconds for VSUGS ($T=150$) and $193$ seconds for Collapsed Gibbs Sampling. 

\subsection{Density estimation for flow cytometry data}

We analyzed the flow cytometry data example, which has been studied thoroughly by \cite{manolopoulou2009selection}. Flow cytometers detect fluorescent reporter markers that typically correspond to specific cell surface or intracellular proteins on individual cells, and can assay millions of such cells in a fluid stream in minutes. These data points are associated with one (or more) components of a Gaussian mixture model (\cite{chan2008statistical}) and are from human peripheral blood cells, with 6 marker measurements each: Forward Scatter (measure of cell size), Side Scatter (Measure of cell granularity), CD4 (marker for helper T cells), IFNg+IL-2 (effector crytokines), CD8 (a marker for cytotoxic T cells), CD3 (marker for all T cells);
that is, the observations are 6 dimensional (the priors are modified to Normal-inverse Wishart, which leads to a similar derivation of the VSUGS algorithm as in Section \ref{sec:dpm_normal}, in this multivariate scenario). Our objective is to compare the performance of VSUGS to SUGS and Collapsed Gibbs Sampling for clustering and density estimation in this multivariate, large data setting.

{\it Data.} The size of the whole data is $50,000$ with $6$ dimensions and \cite{manolopoulou2009selection} state the components of these data are centered closely. In the following simulations, we adopted a Gamma prior for $\alpha$, i.e.~$\alpha\sim\mathcal{G}(1,1)$ for the three approaches.  When considering $\alpha$ as unknown we use the approach to handling uncertainty in $\alpha$ described in \cite{wang} for all algorithms.  The Collapsed Gibbs sampler was run for a 300 iteration burn-in followed by 1000 iterations. This low number is adopted due to the size and complexity of the data; these type of data scenarios are exactly those which motivate the development of SUGS and VSUGS algorithms. For the VSUGS approximation, the truncation value $T$ is set to be $40$ (we did not find significant differences in our results when $T$ is increased or decreased by around 10). We chose the permutation of the order of the data for VSUGS and SUGS as in the previous example.

{\it Results.} We first compared the computation time for the three method with $N=1,000$ data points randomly choose from the whole data set. This process is repeated for $100$ times and we took the average value of the time cost. The analyses through Collapsed Gibbs sampling were completed in approximately $509$ seconds while approximately $8$ seconds and $14$ seconds were required for SUGS and VSUGS respectively. 

\begin{figure}[!h]
\centering
	\includegraphics[width=\textwidth]{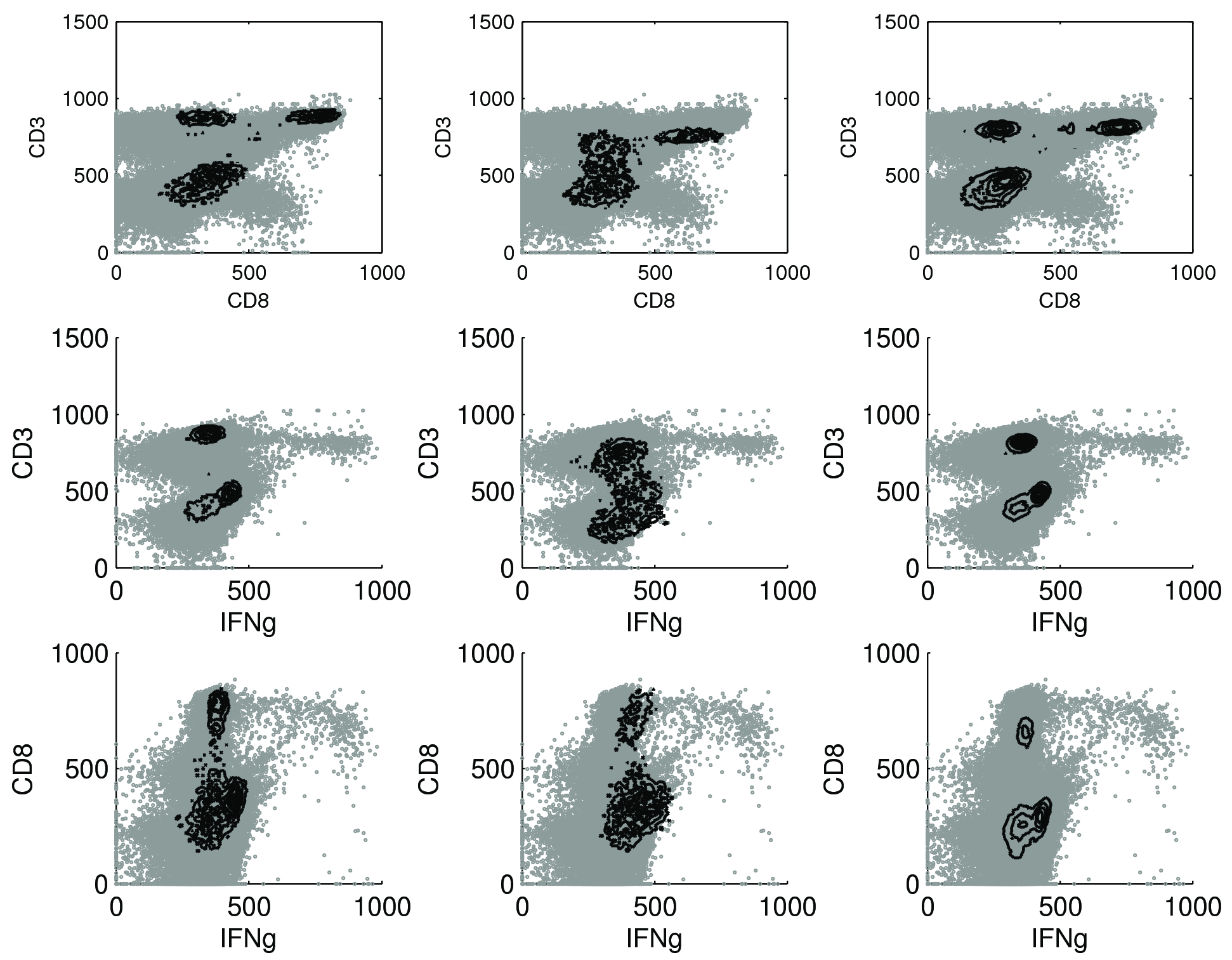}
\caption{Flow Cytometry density estimation examples. Scatter plots of the two-dimensional slices of the multivariate dataset (grey) and contour plots (black) showing density estimates. The contour plot is the estimated distribution through (columns from left to right)  collapsed Gibbs sampling,  SUGS and VSUGS.}
\label{fig:cyto}
\end{figure}

Next, we choose another data sample of $49,000$ data points. We were interested in the performance of all approaches in clustering and density estimation (i.e. the predictive density). The predictive density is calculated on the remaining $1,000$ data points; the Collapsed Gibbs Sampler analysis was repeated $30$ times. The performance of predictive density estimates obtained by the three approaches are shown in Table \ref{tab:flow}. The Collapsed Gibbs sampling method has the greatest predictive ability with VSUGS showing greater predictive power than SUGS. This illustrates that the VSUGS approximation is performing better than SUGS with regard to density estimation and provides an efficient way of detecting and drawing inferences about rare populations in the presence of very large datasets. Figure \ref{fig:cyto} shows that SUGS has difficulty approximating the data density whilst our VSUGS approach better approximates the density estimates by Gibbs Sampling. 

\subsection{SNP Genotyping}

We now turn to our original motivating SNP genotyping example and examined the use of VSUGS and SUGS for a hierarchical Bayesian clustering problem.

\textit{Data.} For our experiments, we considered a genotyping dataset that were considered in a recent comparison study \cite{giannoul}. The study consists of $6$ different individuals, each individual was genotyped three times using the Illumina HumanHap650 genotyping array which produces approximately 650,000 two-dimensional measurements per sample. We normalised the data by taking $\log_2$ transforms and performed quantile normalisation between the two channels to correct for allele-specific biases. 

\textit{Model.} We clustered the data using a three-class Bayesian mixture model:
\begin{align*}
	Y_i | X_i & \sim P(\cdot|X_i), \\
	X_i | w & \sim \mathcal{M}(w_{1:3}),
\end{align*}
where $\mathcal{M}(w_{1:3})$ is the multinomial distribution, we fixed $w_1=w_2=w_3=1/3$ and the class conditional density $P(\cdot|X)$ is given by a Dirichlet Process Mixture of Bivariate Normal Distributions (one DPM for each genotype). We implemented the model using both the SUGS and VSUGS approaches to fit the DPMs.

For comparison, we classified the genotyping data using a standard genotyping tool, GenoSNP \cite{giannoul} which models the class-conditional densities using multivariate Student-$t$ distribution and also performs inference using variational methods. We used majority vote over the three replicates per sample to obtain the \emph{true} genotypes from the GenoSNP genotype calls.

\textit{Results.} Over the $6 \times 3 = 18$ samples, the average concordance of our VSUGS implementation was 99.45\% compared to 98.90\% for the SUGS implementation. Figure \ref{fig:genotyping_example} illustrates genotyping performance for one particular sample. Figure \ref{fig:genotyping_example}(c) indicates that, using genotype calls from GenoSNP as a reference, VSUGS produced the highest concordance with the GenoSNP results across a range of GenoSNP call probability thresholds. 
For the SNPs with discordant genotype calls between GenoSNP and SUGS/VSUGS, we plotted the empirical distribution of the maximum genotype call probabilities for these SNPS. Figure \ref{fig:genotyping_example}(d) shows that for VSUGS the discordant genotype calls were associated with SNPs where the maximum genotype classification probability was around 0.5. With SUGS, discordant calls have probabilities in excess of 0.5.

\section{Summary}\label{sec:summ}

In this paper we have considered VSUGS as a generalization of the SUGS algorithm for fast inference from DPM models. We saw that when the components of the mixture appear to be close in some sense, VSUGS seems to consistently outperform SUGS with regards to density estimation and this improvement is also found by using our variational lower-bound for model selection.
In addition, when $\alpha$ grows, we have found VSUGS performs significantly better, with less computation time. We have found that for real data examples, VSUGS can detect features of the data which SUGS cannot.

In terms of extensions to our work, we are currently considering the development of VSUGS for new models. In particular, we are developing the ideas for hierarchical mixture models and infinite hidden Markov models. These initial experiments suggests that VSUGS can prove to be a very efficient tool for fast, but approximate, inference from a wide class of statistical models.

\section*{Acknowledgements}

We thank Ioanna Manolopoulou for providing codes and data for the flow cytometry example.
The second and fourth authors acknowledge support from the MOE Singapore.

\clearpage

\begin{table}[!h]
\centering
\begin{tabular}{|l|l|l|l|l|l|l|l|l|l|l|l|l}
\hline
$d\mu$ & $N$& 100 & 200 & 300 & 400 & 500 & 600 & 700 & 800 & 900 & 1000\\
\hline
\hline
\multirow{4}{*}{$0.2$} & SUGS & 0.030&0.037&0.049&0.051&0.049&0.045&0.052&0.049&0.052&0.049\\
& VSUGS ($T = 10$) &0.018&0.020&0.023&0.022&0.018&0.015&0.020&0.017&0.021&0.019\\
& VSUGS ($T = 50$) &0.016&0.019&0.021&0.022&0.019&0.018&0.024&0.016&0.020&0.018\\
& VSUGS ($T = 150$) &0.015&0.020&0.021&0.023&0.016&0.017&0.021&0.019&0.020&0.020\\
\hline
\multirow{4}{*}{$0.5$} & SUGS & 0.030&0.037&0.049&0.051&0.049&0.045&0.052&0.049&0.052&0.049\\
& VSUGS ($T = 10$) &0.018&0.020&0.023&0.022&0.018&0.015&0.020&0.017&0.021&0.019\\
& VSUGS ($T = 50$) &0.016&0.019&0.021&0.022&0.019&0.018&0.024&0.016&0.020&0.018\\
& VSUGS ($T = 150$) &0.015&0.020&0.021&0.023&0.016&0.017&0.021&0.019&0.020&0.020\\
\hline
\multirow{4}{*}{$1.0$} & SUGS & 0.030&0.037&0.049&0.051&0.049&0.045&0.052&0.049&0.052&0.049\\
& VSUGS ($T = 10$) &0.018&0.020&0.023&0.022&0.018&0.015&0.020&0.017&0.021&0.019\\
& VSUGS ($T = 50$) &0.016&0.019&0.021&0.022&0.019&0.018&0.024&0.016&0.020&0.018\\
& VSUGS ($T = 150$) &0.015&0.020&0.021&0.023&0.016&0.017&0.021&0.019&0.020&0.020\\
\hline
\multirow{4}{*}{$2.0$} & SUGS & 0.030&0.037&0.049&0.051&0.049&0.045&0.052&0.049&0.052&0.049\\
& VSUGS ($T = 10$) &0.018&0.020&0.023&0.022&0.018&0.015&0.020&0.017&0.021&0.019\\
& VSUGS ($T = 50$) &0.016&0.019&0.021&0.022&0.019&0.018&0.024&0.016&0.020&0.018\\
& VSUGS ($T = 150$) &0.015&0.020&0.021&0.023&0.016&0.017&0.021&0.019&0.020&0.020\\
\hline
\end{tabular}
\caption{Relative error of density estimates of SUGS and VSUGS to Collapsed Gibbs Sampling.}\label{tab:comparison_to_gibbs}
\end{table}

\begin{table}[!h]
\centering
\begin{tabular}{|l|c|c|c|}
\hline
& Gibbs&SUGS&VSUGS\\
\hline
Log Predictive Probability&$-8.2176\times{}10^3$&$-8.8935\times{}10^3$&$-8.4312\times{}10^3$\\
\hline
\end{tabular}
\caption{Log predictive probability on $1,000$ test data points (49,000 training samples) obtained through Collapsed Gibbs sampling, SUGS and VSUGS.}\label{tab:flow}
\end{table}

\begin{figure}[!h]
\centering
	\includegraphics[width=\textwidth]{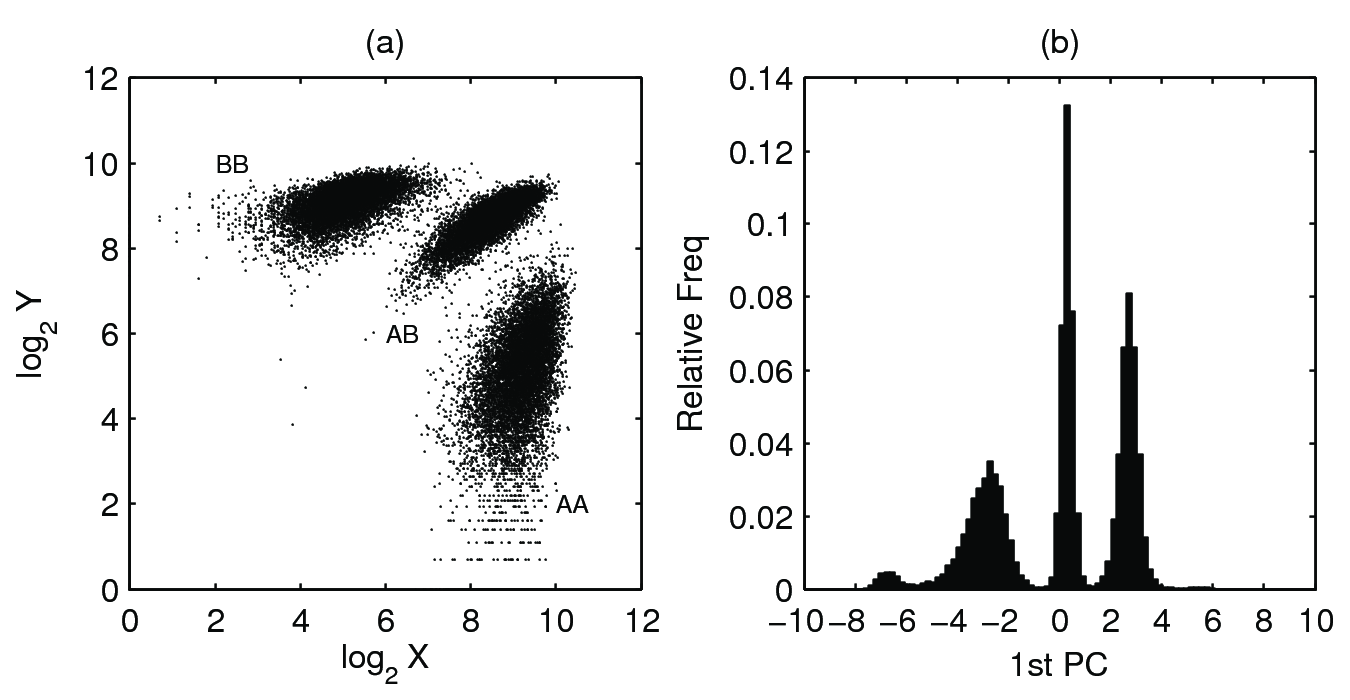}
	\caption{(a) An example SNP genotyping dataset showing three genotype classes $AA$, $AB$ and $BB$ and (b) a transformation using Principal Component Analysis of the same data in (a).}
	\label{fig:genotyping_example}
\end{figure}

\begin{figure}[!h]
\centering
	\includegraphics[width=\textwidth,height=6cm]{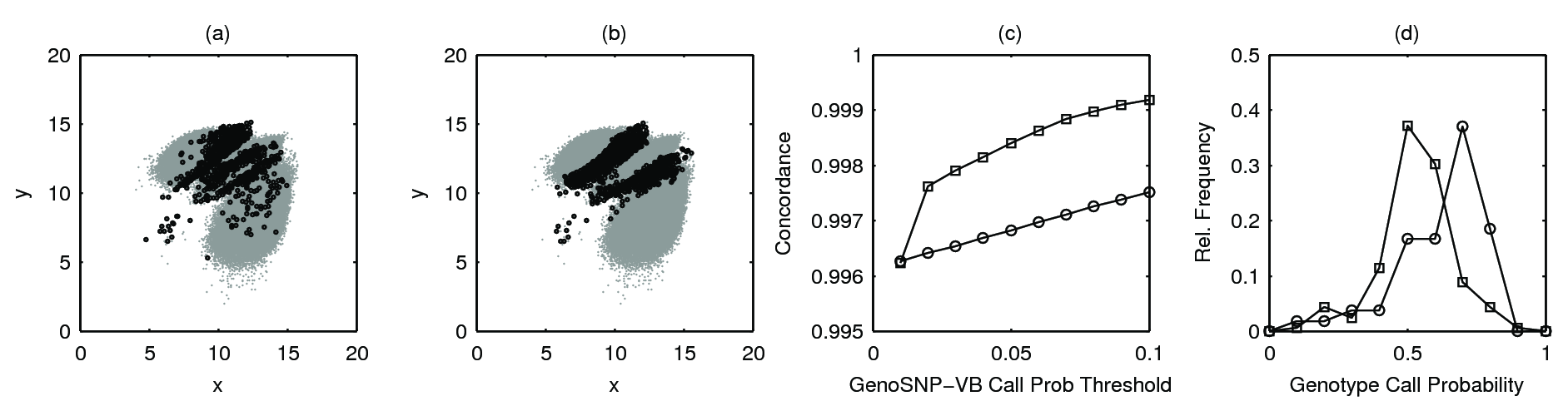}
\caption{Locations of discordant calls between GenoSNP \cite{giannoul} and (a) SUGS, (b) VSUGS. (c) Genotype call concordance between GenoSNP and the (d) distribution of genotype call probabilities for discordant calls ($\circ$) SUGS and ($\square$) VSUGS.  }\label{fig:genotyping_example}
\end{figure}

\end{document}